\documentclass{article}
\usepackage{graphicx}

\begin{document}

\baselineskip=23pt

\begin{center}
\textbf{\Huge Distribution of the extensive Doppler redshift of quasars} %
\vspace{2mm}

\textbf{Yi-Ping Qin$^{1,2}$, Hong-Tao Liu$^{1,3}$, En-Wei Liang $^{1,2}$,
Yun-Ming Dong$^{1,3}$ and Cheng-Yue Su$^{1,3,4}$}

\textbf{$^1$National Astronomical Observatories/Yunnan Observatory, Chinese
Academy of Sciences, P. O. Box 110, Kunming, Yunnan, 650011, P. R. China}

\textbf{$^2$Physics Department, Guangxi University, Nanning, Guangxi,
530004, P. R. China}

\textbf{$^3$ The Graduate School of the Chinese Academy of Sciences}

\textbf{$^4$ Department of Physics, Guangdong Industrial University,
Guangzhou, Guangdong, 510643, P. R. China }
\end{center}

\vspace{4mm}

\begin{center}
\textbf{\Large Abstract}
\end{center}

We make an analysis of the distribution of the extensive Doppler redshift
defined as $\widetilde{z}_{Dopp}=(z_{abs}-z_{em})/(1+z_{em})$ with a sample
of 1317 absorption redshifts available from 401 quasars. The analysis
reveals a bi-peak structure in the distribution, with one component located
at $\widetilde{z}_{Dopp}\simeq 0.00$ and the other at $\widetilde{z}%
_{Dopp}\simeq -0.01$. A study of some possible causes suggests that the
structure can be well interpreted: while the absorbers inside the same
cluster as the quasar concerned could contribute to the first component and
the less populated space between clusters could explain the gap between the
two peaks, the typical distance between clusters could account for the
second component. If the bi-peak structure is true, which needs to be
confirmed by complete samples, one would be able to obtain some useful
cosmological information.

\vspace{2mm}

\begin{flushleft}
{\bf Key words}: cosmology: observations --- galaxies: distances
and redshifts --- quasars: absorption lines --- quasars: emission
lines
\end{flushleft}

\vspace{4mm}

\section{Introduction}

Emission redshifts of quasars are believed to be of cosmological origin.
Absorption redshifts of the objects must be produced by the absorbers in
front of them along the line of sight. In fact, the Lyman-limit systems
(Tytler 1982) are clearly due to intervening gas that is unassociated with
the quasars since they usually appear at redshifts significantly less than
the corresponding emission redshifts. The damped Ly$\alpha $ systems
observed in quasars are thought to arise on sight lines which pass through
galactic disks. Generally, absorption lines of quasars are supposed to arise
in material directly associated with galaxies, such as dark halos.

As emphasized by Peterson (1997), at least two absorption lines are required
for an unambiguous identification and redshift measurement. The most
commonly detected absorption lines are Ly$\alpha $ $\lambda 1216$, CIV $%
\lambda \lambda 1548$, $1551$, and M$_g$II $\lambda 2795$, $2802$. Other
lines that are commonly detected include CII $\lambda 1335$, S$_i$IV $%
\lambda \lambda 1394$, $1403$, M$_g$I $\lambda 2852$, and several UV
resonance lines of F$_e$II.

Obviously, absorption redshifts of quasars can be due to both the
cosmological distance and the Doppler motion of absorbers. Broad absorption
features are detected in the shortward wings of resonance lines of some
quasars (Weymann et al. 1981). The absorption is always at wavelengths
shortward of line center, which indicates that the absorbing gas is flowing
outward from the nucleus, and the high ionization level and high outflow
velocities of the gas strongly suggest that the corresponding absorbers are
closely associated with the nuclear regions. As mentioned in Peterson
(1997), an absorption-line system consists of a number of absorption lines
in a quasar spectrum that are all at very nearly the same redshift $z_{abs}$%
\ and presumably arise in the same absorber. The redshift of these
absorption lines\ will reflect the cosmological distance of the absorbing
cloud rather than that of the quasar which will have emission redshift $%
z_{em}$. It is expected that quasar spectra will show absorption lines
characterized by $z_{abs}<z_{em}$. In most cases this is true (see e.g.,
Junkkarinen, Hewitt, and Burbidge 1991; Hewitt and Burbidge 1993; Qin et al.
2000). Some absorption lines are detected at redshifts slightly larger than $%
z_{em}$\ (see e.g., Weymann et al. 1977; Qin et al. 2000). The $%
z_{abs}>z_{em}$\ phenomenon is thought to be attributable to some
combination of quasars and the absorber peculiar velocities relative to the
Hubble expansion and intrinsic wavelength shifts of the broad absorption
lines relative to their systemic redshifts (Gaskell 1982).

The number of absorption lines we expect along a line of sight to a quasar
is given by $dN(z)=n(z)\sigma (z)dl(z)$, where $n(z)$ is the number density
of absorbers at $z$, $\sigma (z)$ is their cross-section for producing
absorption lines, and $dl(z)$ is an element of proper path length (see
Peterson 1997). This number can reveal evolution in comoving density or
cross-section of absorbers (see e.g., Sargent et al. 1988). Here, we study
another aspect of the absorption feature of quasars, the distribution of the
so-called extensive Doppler redshift which is a relative absorption redshift
defined relative to the corresponding emission redshift. We will try to find
out any cosmological information implied by this distribution.

We present the definition of the extensive Doppler redshift in section 2.
The distribution of the extensive Doppler redshift for a sample of quasars
is presented in section 3. Possible interpretation of the distribution is
discussed in section 4. Conclusions are given in section 5.

\section{Definition of the extensive Doppler redshift}

As mentioned above, absorption redshifts, $z_{abs}$, of quasars reflect both
the Doppler motion and the cosmological distance of the absorbers. When $%
z_{abs}$ is very close to the corresponding emission redshift, $z_{em}$, the
absorber must be associated with the same host galaxy and the difference
between the two redshifts must be produced by the Doppler motion of the
absorber. If $z_{abs}$ is significantly less than the corresponding emission
redshift, the Doppler motion alone would not be able to account for the
difference between the two. Instead, the main part of that difference would
probably be due to distance between the absorber and the host galaxy
concerned (in other words, the absorber would be that associated with
another galaxy). Assuming the absorber and the quasar are associated with
the same host galaxy, the quasar is stationary relative to the cosmological
frame, and the difference between the absorption redshift and the emission
redshift is due to the Doppler motion of the absorber relative to the
quasar, then the Doppler redshift related to the motion is (Kembhavi and
Narlikar 1999)
\begin{equation}
\widetilde{z}_{Dopp}=\frac{z_{abs}-z_{em}}{1+z_{em}}.
\end{equation}
We take this equation as the definition of $\widetilde{z}_{Dopp}$ and apply
it to all absorption redshifts of quasars. The quantity $\widetilde{z}%
_{Dopp} $ is called an extensive Doppler redshift.

Obviously, when $z_{abs}$ is significantly less than $z_{em}$, $\widetilde{z}%
_{Dopp}$ is no more an element reflecting Doppler motions. Instead, it might
become a cosmological quantity. Assuming both $z_{abs}$ and $z_{em}$ arise
from cosmological distances, in terms of the scale factor of the universe $%
R(t)$ we come to
\begin{equation}
\widetilde{z}_{Dopp}=\frac{R(t_{em})}{R(t_{abs})}-1.
\end{equation}
In this situation, $\widetilde{z}_{Dopp}$ becomes in some sense a relative
redshift between the quasar and the absorber.

Generally, for quasars with $z_{abs}<z_{em}$, one can define a relative
redshift $z_{ae}$ between the quasar and the absorber as
\begin{equation}
1+z_{ae}=\frac{R(t_{abs})}{R(t_{em})}=\frac 1{1+\widetilde{z}_{Dopp}},
\end{equation}
with the last equality following from the definitions given above. In terms
of the emission and absorption redshifts the expression of $z_{ae}$ is
\begin{equation}
z_{ae}=\frac{z_{em}-z_{abs}}{1+z_{abs}}.
\end{equation}
When taking $t_{abs}=t_0$ we get $1+z_{ae}=1+z_{em}$ and $1+\widetilde{z}%
_{Dopp}=(1+z_{em})^{-1}$, suggesting that, while $z_{ae}$ is a
conventionally defined redshift, $\widetilde{z}_{Dopp}$ is an inversely
defined one. When $\left| \widetilde{z}_{Dopp}\right| \ll 1$ we come to $%
z_{ae}\simeq -\widetilde{z}_{Dopp}$. It shows that, when $z_{ae}$ and $%
\widetilde{z}_{Dopp}$ act as Doppler redshifts (in the case when $z_{abs}$
and $z_{em}$ are very close, e.g., when the absorber is inside the same host
galaxy of the quasar), $z_{ae}>0$ (while $\widetilde{z}_{Dopp}<0$) suggests
that the absorber moves towards the quasar, and $\widetilde{z}_{Dopp}>0$
(while $z_{ae}<0$) corresponds to the situation when the absorber moves away
from the quasar. [One can find from (4) that, in the case of $z_{abs}\simeq $
$z_{em}$, $z_{ae}$ comes to be a Doppler redshift of the quasar relative to
the absorber.]

\section{Distribution of the redshift}

It is understood that a small value of $\widetilde{z}_{Dopp}$ would reflect
a real Doppler redshift relative to the quasar while a large value of $%
\widetilde{z}_{Dopp}$ would correspond to the case when the absorber is
associated with another galaxy and hence the difference between the
absorption redshift and the corresponding emission redshift would mainly be
due to the cosmological distance between the two galaxies. If we believe
that the majority of the absorbers of quasars are objects associated with
galaxies, then we can expect a gap in the distribution of $\widetilde{z}%
_{Dopp}$ due to the vast distance between galaxies or clusters. If so, one
might probably be able to draw some useful cosmological information from
this gap. In the following we employ a large sample of absorption redshifts
to make the distribution analysis.

In the catalogue of Hewitt and Burbidge (1993), there are 401
quasars with both of their absorption and emission redshifts
available. For these 401 quasars, where emission redshifts range
from $0.158$ to $4.733$, the number of absorption redshifts in
total is 1317. We calculate the extensive Doppler redshift
following (1), where both the absorption and emission redshifts
are taken from the same source, and obtain 1317 extensive Doppler
redshifts.

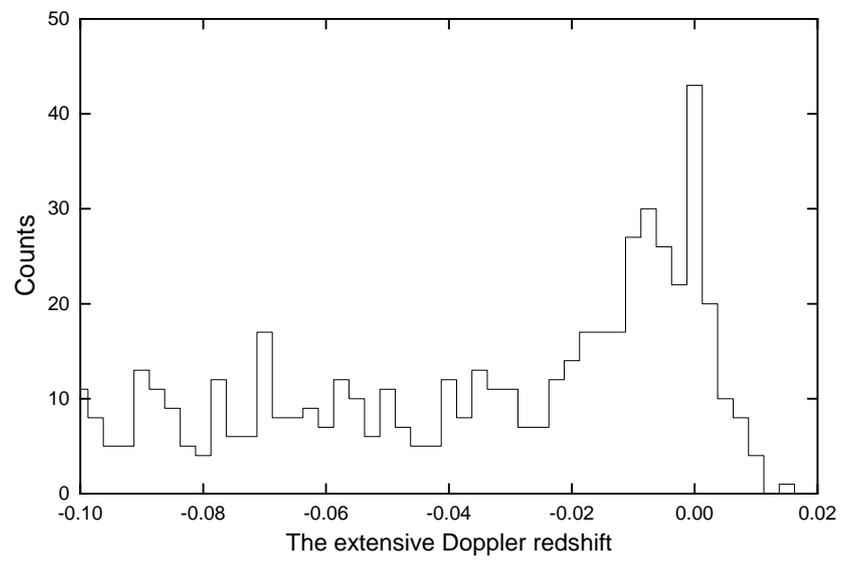
The distribution of the extensive Doppler redshift of the sample is shown in
Fig. 1. As is expected, displayed in the figure, there indeed exists a gap
in the vicinity of the origin. Associated with the gap, there stands a
bi-peak structure with one component located at $\widetilde{z}_{Dopp}\simeq
0.00$ and the other at $\widetilde{z}_{Dopp}\simeq -0.01$.

\section{Cosmological implication of the distribution}

As mentioned above, the extensive Doppler redshift could arise
from the relative velocity between the absorber and the quasar or
the distance between them. Let us investigate if the bi-peak
structure can be produced by the distribution of Doppler
velocities of galaxies. This can straightforwardly be done when a
distribution of Doppler velocities of galaxies is studied.

We employ a large number (544) of proper velocities of galaxies in the Mark
III Catalog of Galaxy Peculiar Velocities (Willick et al. 1997). With these
velocities, we can calculate the corresponding Doppler redshifts by
\begin{equation}
z_{Dopp}=v/c,
\end{equation}
where $c$ is the speed of light. In deriving this formula, the condition of $%
v/c\ll 1$ is assumed. The distribution of $z_{Dopp}$ derived from these
proper velocities is shown in Fig. 2. Displayed in the figure we find that
the distribution of these Doppler redshifts peaks at $z_{Dopp}\simeq
-4.8\times 10^{-4}$. This value is obviously very much smaller in magnitude
than the second component, $\widetilde{z}_{Dopp}\simeq -0.01$, in Fig. 1.
Hence, the assumption that the second component arises from the distribution
of galaxy velocities would not be true. Instead, it might probably be due to
cosmological distances between the absorbers and their corresponding quasars.

It is noticed that the peculiar galaxy velocities are obtained from the
local region, and these might not be the same at high redshifts. However,
while this conjecture would be true, it is still unlikely that the
distribution of galaxy velocities at high redshifts is so different from
that at low redshifts that it would lead to the bi-peak structure while the
latter would not. In addition, if interpreted as the redshift caused by the
proper motion of galaxies, the second component of the bi-peak structure of
the distribution will correspond to a velocity as large as $3000$ km s$^{-1}$%
, suggesting that, besides that at $v_{Dopp}=0$ (corresponding to the first
component), proper velocities of galaxies concentrate at $v_{Dopp}=3000$ km s%
$^{-1}$ as well. This interpretation, either for high or low redshifts, is
obviously not acceptable.

Based on the Friedmann-Robertson-Walker cosmology, one can verify that, in a
flat universe, small intervals of redshift and the radial coordinate can be
related by
\begin{equation}
\Delta z=\frac{H_0R_0\Delta r}c\sqrt{\Omega _m(1+z)^3+1-\Omega _m},
\end{equation}
where $H_0$ is the Hubble constant, $R_0$ is the scale factor and $\Omega _m$
is the parameter of the matter density of the universe at the present epoch.

Assigning $z_{em}-z_{abs}=\Delta z$ and taking $z_{em}=z$ we find from (1)
that $\widetilde{z}_{Dopp}=-\Delta z/(1+z)$. In this way we get
\begin{equation}
\widetilde{z}_{Dopp}=-\frac{H_0R_0\Delta r}c\frac{\sqrt{\Omega
_m(1+z)^3+1-\Omega _m}}{1+z}.
\end{equation}

To calculate the value of $\widetilde{z}_{Dopp}$ with (7) we adopt $%
H_0=100hkms^{-1}Mpc^{-1}$ together with the popular values of $(\Omega
_m,\Omega _\Lambda )=(0.28,0.72)$ (Perlmutter et al. 1999) which correspond
to a flat universe. Let
\begin{equation}
f(z)\equiv \frac{\sqrt{\Omega _m(1+z)^3+1-\Omega _m}}{1+z}.
\end{equation}
For the sample employed, emission redshifts range from $0.158$ to $4.733$.
We find $f(0.158)=0.928$ and $f(4.733)=1.28$ for the adopted parameters.
Therefore, $-1.28H_0R_0\Delta r/c\leq \widetilde{z}_{Dopp}\leq
-0.928H_0R_0\Delta r/c$ is maintained for this sample.

The number density of low-surface brightness galaxies with $23<\mu _0<25$ V
mag/arcsec$^2$ is $N=0.01_{-0.005}^{+0.006}$ galaxies h$_{50}^3$ Mpc$^{-3}$
(Dalcanton et al. 1997); then the typical distance $R_0\Delta r$ between two
neighboring galaxies is about $2.32$ h$^{-1}$ Mpc (note that 1 h$_{50}^3$ Mpc%
$^{-3}$ = 8 h$^3$ Mpc$^{-3}$), which leads to $H_0R_0\Delta r/c\simeq
7.5\times 10^{-4}$. Thus, for emission redshifts ranging from $0.158$ to $%
4.733$, this typical distance would correspond to the extensive Doppler
redshift in the range of $-9.6\times 10^{-4}\leq \widetilde{z}_{Dopp}\leq
-7.0\times 10^{-4}$. The value of $\widetilde{z}_{Dopp}$ obtained here is
far less in magnitude than the second component, $\widetilde{z}_{Dopp}\simeq
-0.01$. If high-surface brightness galaxies are considered, the average
distance between galaxies becomes smaller than $2.32\times h^{-1}$ Mpc, and
then the situation would become worse. Therefore the assumption of the
second component reflecting the average distance of galaxies can be ruled
out.

The density of rich galaxy clusters is about $10^{-5}$ to $10^{-6}$ h$^3$ Mpc%
$^{-3}$ (Bahcall and Cen 1993). Then the average distance between two
neighboring clusters is $\sim 46.4$ to $100$ h$^{-1}$ Mpc. Taking $R_0\Delta
r\simeq 46.4$ h$^{-1}$ Mpc one obtains $H_0R_0\Delta r/c\simeq 0.015$, while
taking $R_0\Delta r\simeq 100$ h$^{-1}$ Mpc leads to $H_0R_0\Delta r/c\simeq
0.032$. For emission redshifts ranging from $0.158$ to $4.733$, they
correspond to the extensive Doppler redshift in the ranges of $-0.019\leq
\widetilde{z}_{Dopp}\leq -0.014$ and $-0.041\leq \widetilde{z}_{Dopp}\leq
-0.030$, respectively. This shows that the value of $\widetilde{z}_{Dopp}$
caused by the typical distance between clusters shares the same order of
magnitude of that of the second component. Hence, while the absorbers inside
the same cluster as the quasar concerned could contribute to the first
component, the typical distance between clusters could account for the
second component. The density of galaxies in the space between clusters must
be less than that within clusters, and this can account for the gap between
the two peaks. Due to this gap and the two peaks, a bi-peak structure would
naturally be formed.

\section{Discussion and conclusions}

In this section, we discuss some issues regarding the reality of the two
peaks and the gap.

If the peak at $\widetilde{z}_{Dopp}\simeq -0.01$ can indeed be accounted
for by the average spacing between clusters, are there many peaks separated
by this spacing in the distribution of the redshift? To find an answer to
this, we make a power spectral analysis and find no periods existing in the
distribution. In fact, the spacing between cosmological sources is not a
linear function of $\widetilde{z}_{Dopp}$ when the spacing is not small
enough. For a flat universe, one has
\begin{equation}
r_{em}-r_{abs}=\frac c{H_0R_0}\int_{1+z_{abs}}^{1+z_{em}}\frac 1{\sqrt{%
\Omega _Mx^3+1-\Omega _M}}dx.
\end{equation}
Applying (1) we come to
\begin{equation}
\frac{H_0R_0(r_{em}-r_{abs})}c=\int_{(1+z_{em})(1+\widetilde{z}%
_{Dopp})}^{1+z_{em}}\frac 1{\sqrt{\Omega _Mx^3+1-\Omega _M}}dx.
\end{equation}
According to (10), it is understandable why we do not find any periods in
the distribution of $\widetilde{z}_{Dopp}$.

Even if no periodicity in the distribution of $\widetilde{z}_{Dopp}$ is to
be expected, should we have seen more than one peak in the distribution,
corresponding to intervening clusters at multiples of the average cluster
distance? Let us recall the fact that the gap, which is expected to
represent the space between galaxies, is not detected. As explained above,
this gap as well as other possible nearby gaps are entirely covered by the
distribution of the proper motions of galaxies. We suspect that there might
be other factors that seal other gaps between clusters. A probable factor
might be the random distribution of clusters that makes the observed
distance between the host galaxy of the quasar and the first absorbing
cluster to vary significantly and the distribution of this distance might
entirely cover the expected gaps. However, no matter how the distribution
is, there will be a lower limit of the concerned distance. (Suppose that
clusters concerned here are not merging. If there are two clusters located
very closely, they would be kept away from each other due to the angular
momentum, otherwise they would be merged. The distance that keeps two
closely located clusters away and not being merged can be regarded as that
lower limit.) It would probably be this lower limit that allows the gap,
which we observe in Fig. 1, to be available.

Note that the quasar sample employed comes from a catalog and is
in no sense complete. Even for all these quasars, the derived
sample consisting of the 1317 absorption redshifts is not complete
at all. Thus, our result is not conclusive.

Because of the nature of the catalogue, the quality of the data is very
heterogeneous. As the typical value of the extensive Doppler redshift we
concerned is as small as $-0.01$, the result might be affected by some poor
data included in our analysis. Contained in our sample, there are some BAL
quasars. Since these objects have absorption lines close in redshift to
emission lines due to outflows related to the quasars, they may obviously
affect the above distribution analysis. With this in mind, we omit 34 BAL
quasars found in the 401 sources and get a subsample containing 1200
extensive Doppler redshifts. The distribution of the extensive Doppler
redshift of this subsample is shown in Fig. 3. Displayed in the figure, we
find that the bi-peak structure is maintained.

Though for different sources of a catalogue, as we what we employ in this
paper, the absorption systems as well as the thresholds for detection of the
systems might be much different, it is still possible to estimate the
measurement errors according to the number of significant figure of the
given value. Here, we simply assume the uncertainty of each absorption
redshift provided in the catalogue is a unit of its last significant figure,
which we call the first uncertainty. In addition, we consider a worse
situation, taking 10 times of the first uncertainty as another uncertainty
which is called the second uncertainty. For example, for an absorption
redshift of 2.5167 presented in the catalogue, we take 0.0001 as its first
uncertainty and 0.001 as its second uncertainty, while for that of 2.98, we
take 0.01 as its first uncertainty and 0.1 as its second uncertainty.

Presented in Chaffee et al. (1988) is a $z_{abs}=1.77642$ system
for MC 1331+170. We find from Table 2 of Chaffee et al. (1988)
that the observed width of the 1328.83 $\AA $ line can be as small
as 0.04 $\AA $. Taking the uncertainty of this line as 0.02 $\AA
$, we get the resultant uncertainty of the absorption redshift as
0.000015. In our sample, the smallest value of the first
uncertainty of the absorption redshift is 0.0001. It seems that
the first uncertainty we defined above is quite reasonable for
many sources.

With the first uncertainty, we get Fig. 4, where, for the sake of
comparison, the curve of Fig. 1 is also plotted. Shown in this figure, the
bi-peak structure is obviously seen. While the position of the first peak
remains unchanged, the range of the position of the second peak spreads
mildly. With the second uncertainty we have Fig. 5. Probably due to the big
uncertainties adopted, the counts spread more randomly in this figure.
However, the bi-peak structure can also be seen (though being less obvious
and shifted slightly, and with less counts for the two peaks). This suggests
that, the main conclusion obtained above, the existence of the bi-peak
structure, is not significantly affected by the measurement error of the
adopted absorption redshifts.

To estimate the significance of the existence of the two peaks as well as
the gap between them, we simply assume that the bi-peak structure is not
true and then check if this assumption is acceptable in terms of statistics.

Suppose the peaks and the gap do not exist according to the distribution of
the probability of events but result from fluctuation. One therefore can
assume a monotonic function of probability in the vicinity of the gap.
Theoretically, one should try all possible monotonic functions and then find
out the one with the largest probability. Here, according to Fig.1, we
simply assume a parabolic function of the distribution of the probability
ranging from $\widetilde{z}_{Dopp}=-0.0250$ to $\widetilde{z}_{Dopp}=0.0025$%
, covering the gap as well as the two peaks. The best fit of the data in
this range yields a value of $\chi ^2=8.40$ (where, the number of data
points is $11$) which corresponds to the following function of the count
density: $C(\widetilde{z}_{Dopp})=-5150000\widetilde{z}_{Dopp}^2+318000%
\widetilde{z}_{Dopp}+14400$, where, the integral of
$C(\widetilde{z}_{Dopp})$ over the range is set to be the same as
the sum of the observed count within the same range. Shown in Fig.
6 are the expected counts of this curve, within the corresponding
bins.

Let us study the statistical significance of the existence of the bi-peak
structure in two different ways. First, we simply examine the goodness of
fit of the above function with the observed counts, which can be realized by
calculating the probability of the $\chi ^2$ obtained above. Here, we take
the function obtained above as the null hypothesis. As the integral of $C(%
\widetilde{z}_{Dopp})$ over the range is set to be the same as the sum of
the observed count within the range and the number of data points is $11$,
the number of degrees of freedom is $10$. The probability of exceeding the
value of $\chi ^2$ obtained above for the given number of degrees of freedom
is only $P\{\chi ^2=8.40,\nu =10\}=0.590$, which indicates that the fit is
poor.

Second, let us consider the probabilities of the occurrence of the three
events: the second peak, the gap, and the first peak. Taking the function
obtained above as the null hypothesis, the probability of the occurrence of
the observed relative frequency of the event within a certain bin in this
range would be available, which is $\int_{\widetilde{z}_{Dopp,1}}^{%
\widetilde{z}_{Dopp,2}}C(\widetilde{z}_{Dopp})d\widetilde{z}_{Dopp}/n$, here
$n=1317$ is the total count concerned. From the best fit function, we get
the probabilities of $p_{peak,2}=0.0233$, $p_{gap}=0.0266$, and $%
p_{peak,1}=0.0282$ for the events of the second peak, the gap, and the first
peak, respectively. One finds from Fig. 1 that the counts of the second
peak, the gap, and the first peak are $n_{peak,2}=36$, $n_{gap}=22$, and $%
n_{peak,1}=45$, respectively. We observe that, the three events are not
independent. Let us redefine the three events by adjusting their relative
frequencies (or the corresponding total counts) and probabilities so that
they are independent. Let the relative frequency of the second peak be $%
n_{peak,2}/n$ and its probability be $p_{peak,2}$. Then, there are $%
n-n_{peak,2}$ counts left. For the gap, while its relative frequency can be
taken as $n_{gap}/(n-n_{peak,2})$, its probability now becomes $%
p_{gap}/(1-p_{peak,2})$. In the same way, for the first peak, while its
relative frequency can be taken as $n_{peak,1}/(n-n_{peak,2}-n_{gap})$, its
probability now becomes $p_{peak,1}/(1-p_{peak,2}-p_{gap})$. The three
events so defined are now independent. The probabilities of the occurrence
of the three independent events can now be calculated by
\begin{equation}
p_{peak,2}^{\prime }\equiv P\{\left| \frac{n_{obs}}n-p_{peak,2}\right|
>\left| \frac{n_{peak,2}}n-p_{peak,2}\right| \},
\end{equation}
\begin{equation}
p_{ap}^{\prime }\equiv P\{\left| \frac{n_{obs}}{n-n_{peak,2}}-\frac{p_{gap}}{%
1-p_{peak,2}}\right| >\left| \frac{n_{gap}}{n-n_{peak,2}}-\frac{p_{gap}}{%
1-p_{peak,2}}\right| \},
\end{equation}
and
\begin{equation}
\begin{array}{c}
p_{peak,1}^{\prime } \\
\equiv P\{\left| \frac{n_{obs}}{n-n_{peak,2}-n_{gap}}-\frac{p_{peak,1}}{%
1-p_{peak,2}-p_{gap}}\right| >\left| \frac{n_{peak,1}}{n-n_{peak,2}-n_{gap}}-%
\frac{p_{peak,1}}{1-p_{peak,2}-p_{gap}}\right| \},
\end{array}
\end{equation}
respectively, where $n_{obs}$ denotes the observed count within the bin
concerned.

It is known that, for a certain probability $p$, when the total count $%
n\rightarrow \infty $, the distribution of the observed relative frequency $%
n_{obs}/n$ would approach that of gauss, with its mean being $p$ and its
variance being $p(1-p)/n$. Calculating with the gauss distribution, we
obtain the probabilities for the occurring of the three independent events ($%
E_1\equiv \{\left| \frac{n_{obs}}n-p_{peak,2}\right| >\left| \frac{n_{peak,2}%
}n-p_{peak,2}\right| \}$, $E_2\equiv \{\left| \frac{n_{obs}}{n-n_{peak,2}}-%
\frac{p_{gap}}{1-p_{peak,2}}\right| >\left| \frac{n_{gap}}{n-n_{peak,2}}-%
\frac{p_{gap}}{1-p_{peak,2}}\right| \}$ and $E_3\equiv \{\left| \frac{n_{obs}%
}{n-n_{peak,2}-n_{gap}}-\frac{p_{peak,1}}{1-p_{peak,2}-p_{gap}}\right|
>\left| \frac{n_{peak,1}}{n-n_{peak,2}-n_{gap}}-\frac{p_{peak,1}}{%
1-p_{peak,2}-p_{gap}}\right| \}$): $p_{peak,2}^{\prime }=0.324$ for $E_1$, $%
p_{gap}^{\prime }=0.0261$ for $E_2$, and $p_{peak,1}^{\prime }=0.199$ for $%
E_3$, respectively. The probability of none of the three events occurring is
$(1-p_{peak,2}^{\prime })(1-p_{gap}^{\prime })(1-p_{peak,1}^{\prime })=0.527$
(the events of $E_1$, $E_2$ and $E_3$ not occurring). The probabilities of
one of the three events occurring but the other two not occurring are $%
p_{peak,2}^{\prime }(1-p_{gap}^{\prime })(1-p_{peak,1}^{\prime })=0.253$
(the event of $E_1$ occurring but the events of $E_2$ and $E_3$ not
occurring), $(1-p_{peak,2}^{\prime })p_{gap}^{\prime }(1-p_{peak,1}^{\prime
})=0.0141$ (the event of $E_2$ occurring but the events of $E_1$ and $E_3$
not occurring), and $(1-p_{peak,2}^{\prime })(1-p_{gap}^{\prime
})p_{peak,1}^{\prime }=0.131$ (the event of $E_3$ occurring but the events
of $E_1$ and $E_2$ not occurring). The probabilities of two of the three
events occurring but the other one not occurring are $p_{peak,2}^{\prime
}p_{gap}^{\prime }(1-p_{peak,1}^{\prime })=0.00677$ (the events of $E_1$ and
$E_2$ occurring but the event of $E_3$ not occurring), $p_{peak,2}^{\prime
}(1-p_{gap}^{\prime })p_{peak,1}^{\prime }=0.0628$ (the events of $E_1$ and $%
E_3$ occurring but the event of $E_2$ not occurring), and $%
(1-p_{peak,2}^{\prime })p_{gap}^{\prime }p_{peak,1}^{\prime }=0.00351$ (the
events of $E_2$ and $E_3$ occurring but the event of $E_1$ not occurring).
The probability of all the three events occurring is $p_{peak,2}^{\prime
}p_{gap}^{\prime }p_{peak,1}^{\prime }=0.00168$ (the events of $E_1$, $E_2$
and $E_3$ occurring). The observational fact is that all the three events ($%
E_1$, $E_2$ and $E_3$) occur. Thus, the null hypothesis is not
acceptable. In views of statistics, the existence of the bi-peak
structure is significant.

To find out if the bi-peak structure is stable, let us examine the
distribution of the extensive Doppler redshift with subsamples. One should
notice that the largest value of counts in Fig. 1 is $45$. This suggests
that subsamples selected should not be much smaller than the total sample.
We hence select a subsample with 200 sources out of the 401 source sample.
The selection is made with the method of random sampling and we obtain a 200
source sample with 655 absorption redshifts. We therefore have 655 extensive
Doppler redshifts. The distribution of $\widetilde{z}_{Dopp}$ of this
subsample is shown in Fig. 7. We find from the figure that the bi-peak
structure is still observable. In the same way, we get the best fit
parabolic function of the distribution of the probability, ranging from $%
\widetilde{z}_{Dopp}=-0.0250$ to $\widetilde{z}_{Dopp}=0.0025$, as $C(%
\widetilde{z}_{Dopp})=-8980000\widetilde{z}_{Dopp}^2-1.02\widetilde{z}%
_{Dopp}+6610$. The fit yields $\chi ^2=9.54$ (where, the number of data
points is $11$). For the same reason, the number of degrees of freedom is
taken to be $10$. The probability of exceeding the value of $\chi ^2$
obtained above for the given number of degrees of freedom is only $P\{\chi
^2=9.54,\nu =10\}=0.482$, which indicates that the fit is poor. Also in the
same way, we redefine the three events (the second peak, the gap, and the
first peak) to make them independent. Then we get the probabilities of the
three independent events ($p_{peak,2}^{\prime }=0.262$, $p_{gap}^{\prime
}=0.0187$, $p_{peak,1}^{\prime }=0.397$) and the product of them, which is $%
P=0.00195$. It shows that the existence of the bi-peak structure
is significant as well with this subsample. When selecting a
smaller subsample (a 100 source sample with 345 absorption
redshifts), we find that the bi-peak structure is poorly
detectable. We come to the conclusion that the bi-peak structure
is detectable as long as the sample employed is large enough.

Besides the distribution of the extensive Doppler redshift, can one find any
other signs of the clustering in the distribution of other redshifts? As the
extensive Doppler redshift is defined with both absorption and emission
redshifts, we wonder if the distribution of absorption or emission redshifts
individually show some signs of clustering. Shown in Fig. 8 are the
distributions of the absorption and emission redshifts of the 401 source
sample. We do not find any signs of the clustering in this figure. In fact,
any peaks shown in the two distributions would not be able to reveal a
clustering. For example, for a peak appearing in the distribution of
absorption redshifts, only a fraction of all the redshifts within the bin,
which is associated with slightly larger emission redshifts, can be
attributed to the second peak in Fig. 1, and the two distributions
themselves (or a combination of them) cannot tell how large this fraction is.

As mentioned above, the number of absorption lines one expects along a line
of sight to a quasar is given by the product of the number density of
absorbers, the absorption cross-section and the path length. Absorber
properties are expected to change with redshift and one can study the
development of the properties with this number. However, the changes of
either the number density or the absorption cross-section are unlikely to
ensure that the less populated space between clusters could provide
sufficient absorbers that can affect the bi-peak structure of the
distribution, while possibly, they could affect the distribution of large $%
\widetilde{z}_{Dopp}$ which corresponds to large distances. Compared with
this approach, the distribution analysis adopted in this paper refers to a
different quantity, the extensive Doppler redshift (which is a relative
absorption redshift defined relative to the corresponding emission
redshift). The analysis is also able to draw some useful cosmological
information. For example, when several samples with different emission
redshifts are available, one can determine the development of the distance
between clusters and then can constrain the cosmological model.

\vspace{6mm}

\begin{center}
{\Large \textbf{Acknowledgments}}\\
\end{center}

It is our great pleasure to thank Dr. A. Kembhavi for his helpful
suggestions. This work was supported by the Special Funds for Major State
Basic Research Projects (``973'') and National Natural Science Foundation of
China (No. 10273019).

\vspace{6mm}

\begin{center}
\textbf{REFERENCES}
\end{center}

\begin{verse}
Bahcall, N.A. and Cen, R. 1993, ApJ, 407, L49

Chaffee, Jr., F. H., Black, J. H., and Foltz, C. B. 1988, ApJ, 335, 584

Dalcanton, J. J. et al, 1997, ApJ, 114, 635

Gaskell, C. M. 1982, ApJ, 263, 79

Hewitt, A. and Burbidge, G. 1993, ApJS, 87, 451

Junkkarinen, V., Hewitt, A. and Burbidge, G. 1991, ApJS, 77, 203

Kembhavi, A. K. and Narlikar, J. V. 1999, in Quasars and Active Galactic
Nuclei, (Cambridge: Cambridge University Press)

Perlmutter, S., et al. 1999, ApJ, 517, 565

Peterson, B. M. 1997, in An Introduction to Active Galactic Nuclei,
(Cambridge: Cambridge University Press)

Qin, Y.-P., Xie, G.-Z., Zheng, X.-T., and Liang, E.-W. 2000, astro-ph/0005006

Sargent, W. L. W., Boksenberg, A., and Steidel, C. C. 1988, ApJS, 68, 539

Tytler, D. 1982, Nature, 298, 427

Weymann, R. J., Carswell, R. F., and Smith, M. G. 1981, ARA\&A, 19, 41

Weymann, R. J., Willams, R. E., Beaver, E. A., and Miller, J. S. 1977, ApJ,
213, 619

Willick, J. A., Courteau, S., Faber, S. M., et al. 1997, ApJS, 109, 333
\end{verse}

\end{document}